\documentclass[notitlepage,twocolumn,prl,tightenlines,nofootinbib,superscriptaddress,showpacs]{revtex4}

\usepackage{amsmath,amssymb,amsfonts,latexsym,MnSymbol}
\usepackage{bbm}
\usepackage[mathcal]{euscript}
\usepackage{graphicx,epsfig}
\usepackage{color}
\usepackage[activate=normal]{pdfcprot}  

\newcommand{\eccm}{EC2M, UMR Gulliver 7083 CNRS, ESPCI ParisTech,
PSL Research University, 10 rue Vauquelin, 75005 Paris, France}
\newcommand{\sphynx}{SPHYNX/SPEC, CEA, CNRS, Universit\'e Paris-Saclay, F-91191, Gif-sur-Yvette, France}
\newcommand{\fast}{Laboratoire FAST, Universit\'e Paris-Sud, CNRS, Universit\'e Paris-Saclay, F-91405, Orsay, France}

\newcommand{\be}{\begin{equation}}
\newcommand{\ee}{\end{equation}}
\newcommand{\ben}{\begin{equation*}}
\newcommand{\een}{\end{equation*}}
\newcommand{\ba}{\begin{eqnarray}}
\newcommand{\ea}{\end{eqnarray}}

\newcommand{\bfr}{\mathbf{r}}

\begin{document}
\graphicspath{{./figures/}}

\title{Experimental evidences of the Gardner phase in a granular glass}

\author{A. Seguin}
\affiliation{\sphynx}
\affiliation{\fast}
\author{O. Dauchot}
\affiliation{\eccm}

\date{\today}

\begin{abstract}
Analyzing the dynamics of a vibrated bi-dimensional packing of bidisperse granular discs below jamming, we provide evidences of a Gardner phase deep into the glass phase. To do so we perform several independent compression cycles \emph{within the same glass} and show that the particles select different average vibrational positions at each cycle, while the neighborhood structure remains unchanged. We compute the mean square displacement as a function of the packing fraction and compare it with the average separation between the cages obtained for different compression cycles. Our results are fully compatible with recent numerical observations obtained for a mean field model of glass as well as for hard spheres in finite dimension. We also characterize the distribution of the cage order parameters. Here we note several differences from the numerical results, which could be attributed to activated processes and cage heterogeneities.
\vspace{5mm}
\end{abstract}

\maketitle
The constituent particles of a glass are caged by their neighbors and thus cannot relax density fluctuations~\cite{Cavagna:2009tx}. This is also true for hard particles under compression~\cite{krzakala-2007, Mezard:2012he}. The associated slowing down of the dynamics is related to a complex free energy landscape with multiple glass states~\cite{Goldstein1969,debenedetti2001sla}. This picture is however too simple to describe the complex aging properties of glasses~\cite{Goldstein:2011cv,Fragiadakis:2012jo}, as well as the observation of dynamical heterogeneities in low temperature glasses~\cite{VollmayrLee:2002eb,ballesta2008udd}. Furthermore, when compressing hard particles to infinite pressure, a geometric transition takes place, the jamming transition, at which the dynamics is fully arrested and the particles are mechanically equilibrated~\cite{OHern:2003gu,zamponi2008mft,vanHecke:2009go,Liu:2010vh}. This transition exhibits critical scalings, which characterize the marginal stability of the glass on approaching jamming and are associated with very soft, slow and delocalized excitation modes~\cite{wyartPRE,brito2007hdm,Brito:2009ed,Chen:2010th}. Such features can also not be captured within the above simple landscape picture. 

\begin{figure}[t!] 
\center
\vspace{0cm}
\includegraphics[width=\columnwidth]{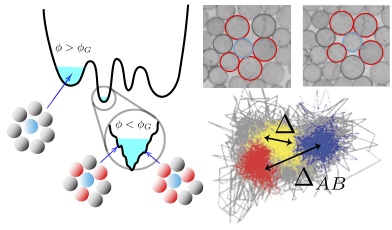}
\vspace{-0.5cm}
\caption{{\bf The Gardner transition:} {\bf Left:} A glass meta basin breaks into a hierarchy of marginally stable sub-basins (freely inspired from~\cite{Charbonneau:1kk}). In the glass phase $(\phi_g<\phi<\phi_G)$ the particle (in blue) is caged by its neighbors; those of which establish contact at jamming $(\phi_J)$ are not selected yet. In the Gardner phase $(\phi>\phi_G)$, each of the sub-basin eventually corresponds to one structure of contact network (red neighbors). {\bf Right-top :} Experimental realization of this scenario in a bi-disperse system of discs for two independent compression up to $\phi_J$ within the same glass state. {\bf Right-bottom :} In the Gardner phase, below jamming $(\phi_G<\phi<\phi_J)$, successive compressions starting from the same glass state, where particles vibrate in a large (grey) cage, will lead to a different caging location in a smaller cage (blue-red-yellow). While $\Delta(\phi)$, the cage size, decreases with $\phi$, $\Delta_{AB}$, the typical distance between the cages obtained for different compression plateaus at the cage size of the transitional packing fraction, $\phi_G$.}
\label{fig:context}
\vspace{-0.5cm}
\end{figure}

It was recently shown theoretically that the hard sphere glass in infinite dimension undergoes a Gardner transition~\cite{Gardner:1985yq}, at which the glass basin breaks into a hierarchy of marginally stable sub-basins (see fig.~\ref{fig:context}). The associated structure of the free energy landscape is necessary to capture the critical scalings of the jamming transition~\cite{Charbonneau:1kk}. Later, the Gardner transition was detected numerically in a mean-field glass model~\cite{Charbonneau:2015ig}. The critical properties of jamming are independent of the spatial dimension~\cite{Skoge:2006go,Goodrich:2012tl,Charbonneau:2012fl}, and one expects that the above findings apply in finite dimension. This was very recently confirmed in simulations of 2d and 3d hard sphere (HS) glasses~\cite{Berthier:2015db}. 
Yet it remains to be observed experimentally: in practice, finite-size and time effects, activation processes and aging could very well hinder the transition.

In this letter we bring the first direct experimental evidences of the Gardner phase, taking advantage of a well controlled granular experiment, which has already proven to successfully probe the vicinity of the jamming transition in a bi-dimensional granular glass former~\cite{Lechenault:2008tw,Lechenault:2010cl,Brito:2010kc,Candelier:2009kb,Coulais:2014bn,Coulais:2012vj,Berthier:2011bk}. 
More precisely, following the protocol suggested for the numerical detection of the Gardner transition~\cite{Charbonneau:2015ig,Berthier:2015db}, taking place at $\phi_G$, we perform independent compressions of the same glass and show that for large enough compression, the final state differs from one compression to another. To do so we compare the average cage size within one state, $\Delta$, and the average distance separating the cages of the same particles across successive compression cycles, $\Delta_{AB}$. While for $\phi<\phi_G$,  $\Delta_{AB}$ decreases like $\Delta$, it plateaus to a constant value equal to $\Delta(\phi_G)$, when $\phi>\phi_G$. 
Our results are in perfect agreement with the numerical observations obtained within the Mari-Kurchan (MK) mean field model~\cite{mari09,Mari:2011jg,Charbonneau:2015ig} and the HS systems~\cite{Berthier:2015db} and sign the entrance into the Gardner phase. We further characterize the fluctuations of the cage sizes $\Delta$ and inter-cycle cage distances $\Delta_{AB}$ and report some differences from the mean field and HS cases. We finally discuss the possible origin of these differences as well as the importance of the Gardner phase regarding the interpretation of experimental results obtained in former studies.

{\it --- Theoretical Context ---}
Compressing a liquid and avoiding crystallization -- for instance using poly-disperse systems -- one ends up in a glass state. The structure is frozen and particles cannot exchange neighbors anymore. The location of the glass transition $\phi_g$ depends on the compression rate and for each realization a different glass state is selected. Further compressing this glass, the pressure increases until it diverges when particles come into contact and reach mechanical equilibrium at $\phi=\phi_J$; the glass is jammed. In between sits the Gardner transition~\cite{Gardner:1985yq} of that glass.
When $\phi>\phi_G$, the glass meta-basin breaks into a hierarchy of marginally stable sub-basins, the bottom of which correspond to different structures of the contact network (see fig.~\ref{fig:context}). Measuring the similarity of the contact network between jammed structures obtained after independent compressions, is one way of testing the existence of the marginal phase~\cite{Charbonneau:1kk}. Alternatively, one can follow the caged dynamics~\cite{Charbonneau:2015ig,Berthier:2015db}. Since the detection of contacts in experiments is always prone to some arbitrariness, we will follow this second path. 

Starting from a glass state at a packing fraction $\phi_0>\phi_g$, and realizing independent compressions up to a packing fraction $\phi$, one should follow particle trajectories $\bfr^k_i(t)$ in the final state k and compute (i) for each compression the mean square displacement (MSD) $\Delta^k(t,\tau)$ and (ii) for each pair of compressions, the "mean square distance" $\Delta^{kk'}(t)$ between the two compressed states $(k,k')$:
\ba
\label{eq:Delta}
\Delta^k(t,\tau) &=& \frac{1}{N}\sum_{i=1}^N \left| \bfr^k_i(t+\tau) - \bfr^k_i(t) \right|^2 \\
\label{eq:DeltaAB}
\Delta^{k,k'}(t) &=& \frac{1}{N}\sum_{i=1}^N \left| \bfr^k_i(t) - \bfr^{k'}_i(t) \right|^2.
\ea
Averaging over compressions and thermal samples, one obtains $\Delta(t,\tau) = \overline{\left<\Delta^k(t,\tau) \right>}$ and $\Delta_{AB}(t) = \overline{\left< \Delta^{k,k'}(t) \right>}$. For $\phi<\phi_G$, if the glass is well equilibrated (i.e at large enough $t$), $\Delta_{AB}(t) = \lim_{\tau \rightarrow \infty} \Delta(t,\tau)$. On the contrary for $\phi>\phi_G$, equilibrium is never reached and $\Delta_{AB}(t)$ remains larger than $\Delta(t,\tau)$, even at large $t$. The large $\tau$ behavior of $\delta\Delta(t,\tau) = \Delta_{AB}(t) - \Delta(t,\tau)$ is thus a good dynamical order parameter of the Gardner transition. One sees however that long time limits have to be considered and that aging in the Gardner phase significantly complicates the analysis (see~\cite{Charbonneau:2015ig,Berthier:2015db} for a more complete discussion). In finite dimension, the situation is even less clear. The cages are heterogeneous, and activated dynamics will prevent the configurations at $\phi_0$ from constraining the dynamics at arbitrarily long times so that the glass itself is never fully equilibrated. Moving to experiments, one has to deal with finite time issues, as in simulations, but also with specifically experimental constraints such as limited spatial resolution and possible artifacts such as slow but persistent convection currents~\cite{Coulais:2014un}.

{\it --- Experimental implementation ---} 
The experimental setup already described elsewhere~\cite{Lechenault:2008tw} consists in a monolayer of $8500$ bi-disperse ($0.44\%$ large, $0.56\%$ small) photo-elastic discs of stiffness $\kappa = 1660 N.m^{-1}$, with diameters $d_{s} = 4/5 \, d_{l} = 4\pm0.01 {\rm mm}$. The discs are laid out on a horizontal glass plate vibrated horizontally with frequency $f=10{\rm Hz}$ and amplitude $A=10{\rm mm}$ and confined in a cell fixed in the laboratory frame. The packing fraction, $\phi$, can be varied by tiny amounts ($\delta\phi/\phi \sim 5.10^{-4}$). The stroboscopic motion of a set of 1600 grains in the center of the sample is tracked by a CCD camera synchronized with the plate. The position of the grains is detected with an accuracy of $10^{-2} d_{s}$. In the following, lengths are measured in $d_{s}$ units and time in cycle units. 

Starting from a low packing fraction $\phi$, we gradually compress the system until it reaches a highly jammed state following the same protocol as in~\cite{Coulais:2014un}. Then we stepwise decrease the volume fraction until $\phi_J$, where (i) the pressure measured at the wall in the absence of vibration falls to zero and (ii) the contacts observed through cross-polarizers disappear, signaling the un-jamming transition. The precise value of $\phi_J$ varies with each realization of the preparation protocol. Rescaling several samples using different values of $\phi_J$ would require a perfect control on the value of the packing fraction where the glass reach the liquid state, which we cannot guarantee experimentally. We thus consider a unique sample for which $\phi_J = 0.8236 \pm 0.0003$. Then we gently decompress the system further down to $\phi_0 = 0.8185$, where we check that most of the discs keep the same neighbors (less than $0.5\%$ broken links during $1000$ vibration cycle), ensuring that the system remains in the same glass state. We then perform 10 sets of compression cycles, from $\phi_0$ to $\phi \in [\phi_0, \phi_J]$, each cycle containing 10 compressions. The compression cycles are rapid quenches separated by 1000 vibration cycles.  This number of cycles spent in each compressed state has been chosen so that it is large enough to define cages and small enough as compared to the relaxation time of the instantaneous contact network, which was found to be of the order of 5000 to 10000 cycles in past studies~\cite{Coulais:2014bn,Coulais:2014un}.

\begin{figure}[t] 
\center
\includegraphics[width=0.85\columnwidth]{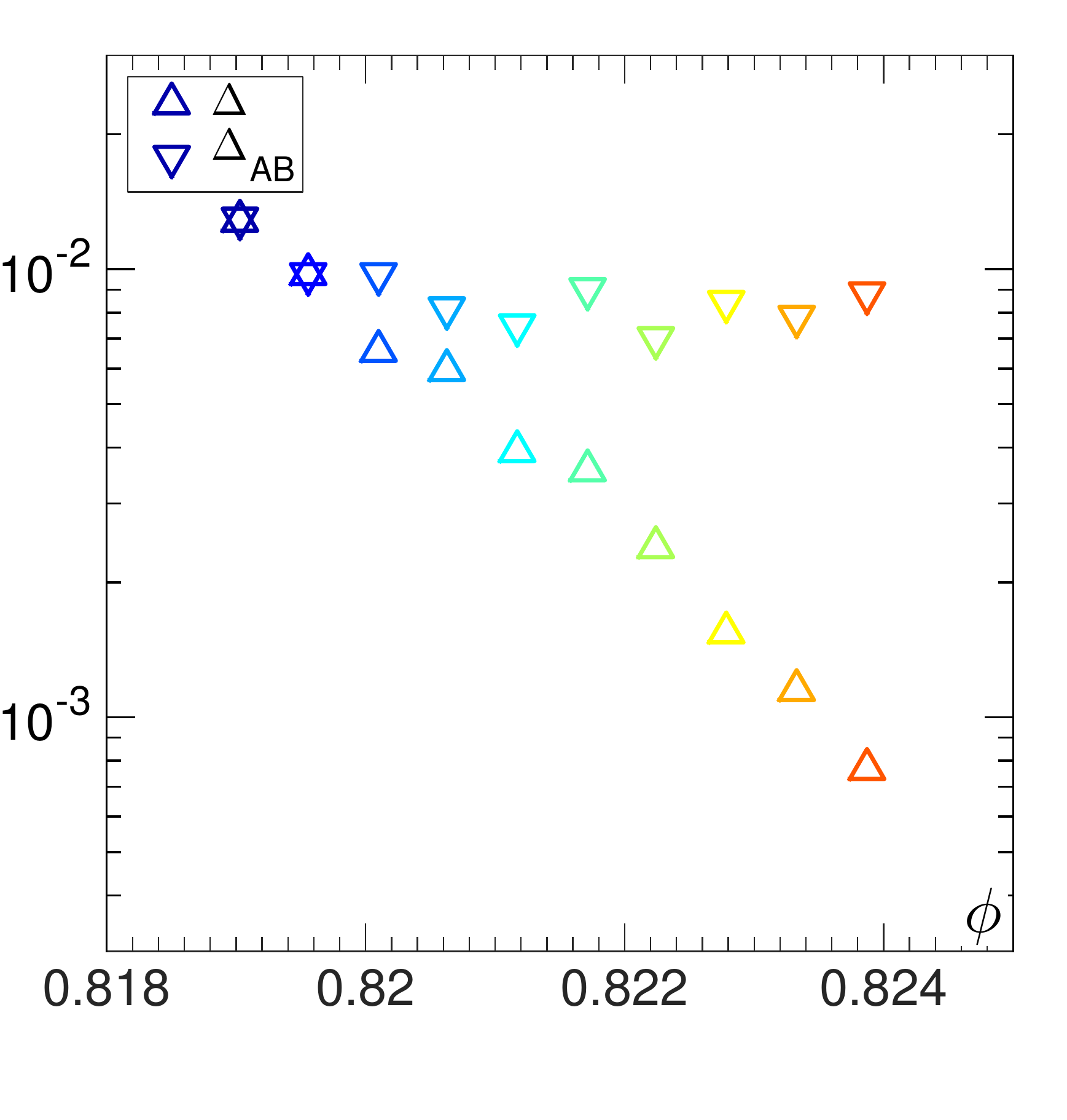}
\vspace{-0.7cm}
\caption{{\bf Cage order parameters.} Mean square displacement at long time $\Delta(\tau_0)$ and mean square distance between cages $\Delta_{AB}$ averaged over $200$ time steps and $10$ compression cycles from $\phi_0 = 0.8185$ to $\phi$ as a function of $\phi$.}
\label{fig:deltas}
\vspace{-0.5cm}
\end{figure}

Vibrated granular matter is prone to develop convection as soon as it un-jams~\cite{Coulais:2014un}. This convection can be very small, but it accumulates over time and removing it is always a challenge.  In order to eliminate spurious effects in the measurements of the displacements, we analyze the dynamics of each particles in the reference frame of the center of mass of its neighbors. 
Because we are interested in individual displacements -- although they result from collective effects --, this technique has the advantage of not subtracting an overall convection field, which, locally, can be large as compared to the intrinsic displacement. We thus redefine the particle position as $\bfr_i \rightarrow \bfr_i - \bfr^\Omega_i$, where $\bfr^\Omega_i = 1/n_i \sum_{j} \bfr_j$, where the sum runs over the neighbors of particle $i$ and $n_i$ is the number of them. We are then in position to compute $\Delta^k(t,\tau)$ and $\Delta^{kk'}(t)$ as defined by equations~(\ref{eq:Delta}) and~(\ref{eq:DeltaAB}). We average over compressions and replace the average over samples by temporal averaging to obtain $\Delta(\tau)$ and $\Delta_{AB}$. Doing so we assume aging to be negligible on the $200$ time steps time window used here. 

Figure~\ref{fig:deltas} displays $\Delta = \Delta(\tau_0=500)$ and $\Delta_{AB}$ as a function of $\phi$. It is the first and main experimental result of the present study. The typical MSD  within cages $\Delta(\phi)$ steadily decreases with the packing fraction. By contrast, $\Delta_{AB}(\phi)$ first decreases like $\Delta(\phi)$, but then plateaus at values of the order of $10^{-2}$ for $\phi\gtrsim0.820$. This unambiguously demonstrates that there is a regime at large $\phi$, before jamming, in which several cage configurations are separated by an average distance larger than the cage size. Also, the fact that $\Delta$ present no sign of a transition shows that the glass meta-basin is not suddenly broken in sub-basins corresponding to a well defined smaller cage size, but rather in a sub-basins with distributed cage sizes. The system has entered the Gardner phase. We estimate that the Gardner transition takes place at $\phi_G\simeq 0.820$. Very remarkably, although the present system is frictional and the jamming packing fractions are lower than for thermal hard discs, our observations very well match the numerical ones~\cite{Charbonneau:2015ig,Berthier:2015db}. Both the plateau value of $10^{-2}$, the ratio $\phi_G/\phi_J \simeq 0.996$ and $1/(\phi_J-\phi_G)\simeq 3.10^2$, a naive estimation of the pressure, are in the correct range. Extrapolating $\Delta$ to $10^{-1}$, the typical MSD values reported for experimental and numerical glass transition, we find $\phi_g = 0.815$, in agreement with previously reported values for the glass transition in 2d systems of bi-disperse hard grains~\cite{abate2006aja,Candelier:2010vk}. 
 
\begin{figure}[t] 
\center
\includegraphics[width=0.49\columnwidth,height=0.49\columnwidth]{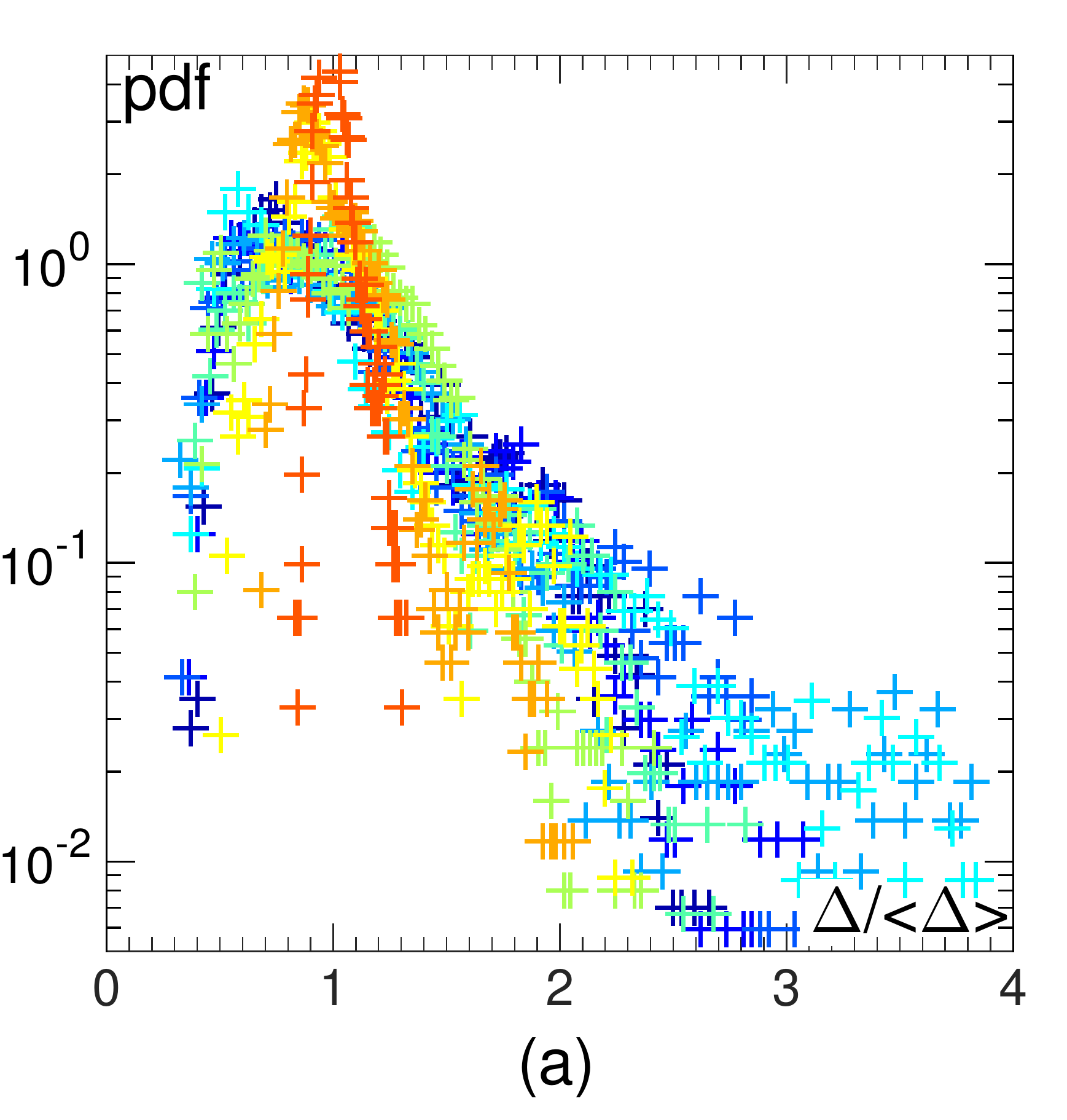}
\includegraphics[width=0.49\columnwidth,height=0.49\columnwidth]{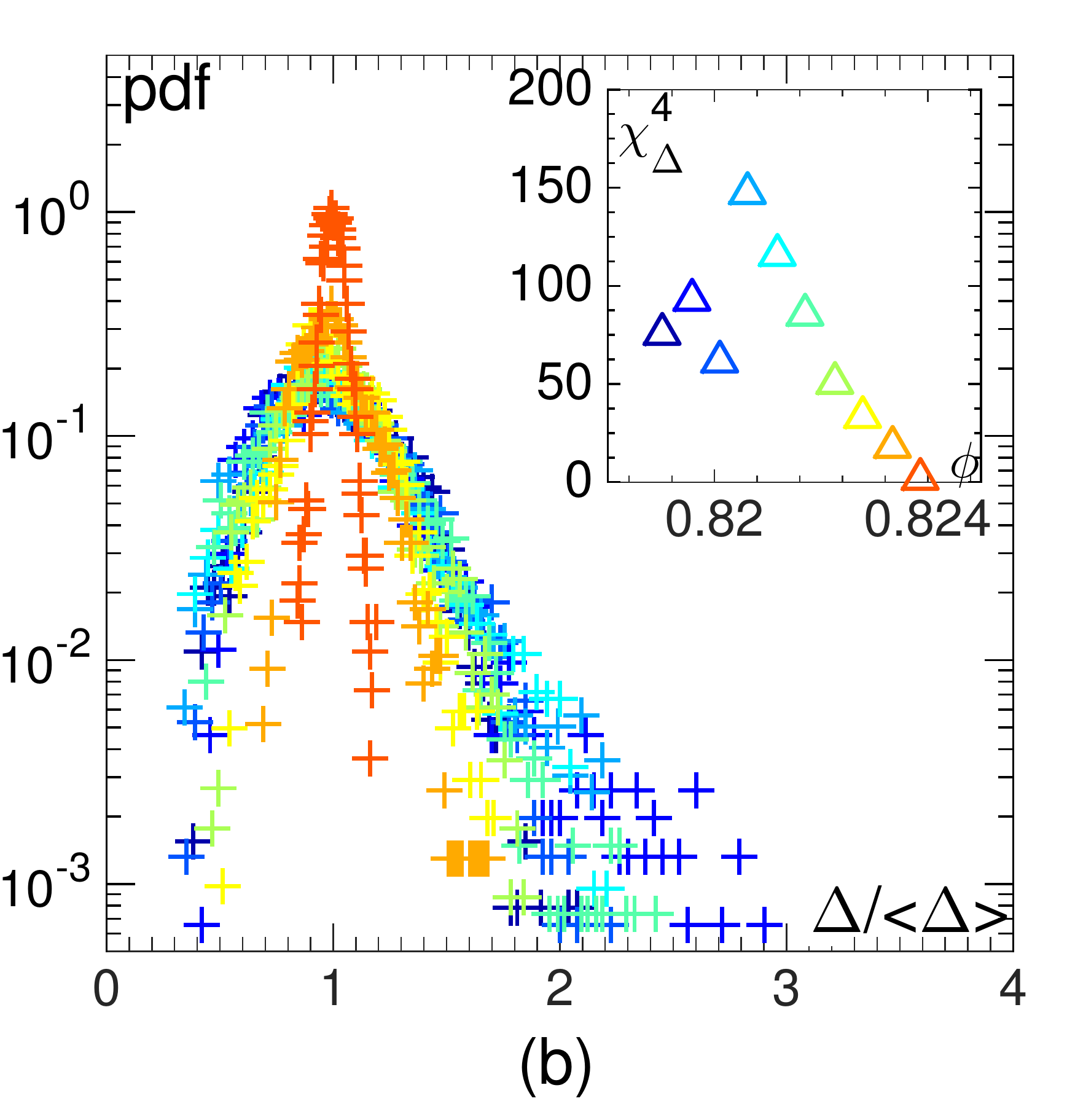}\\
\includegraphics[width=0.49\columnwidth,height=0.49\columnwidth]{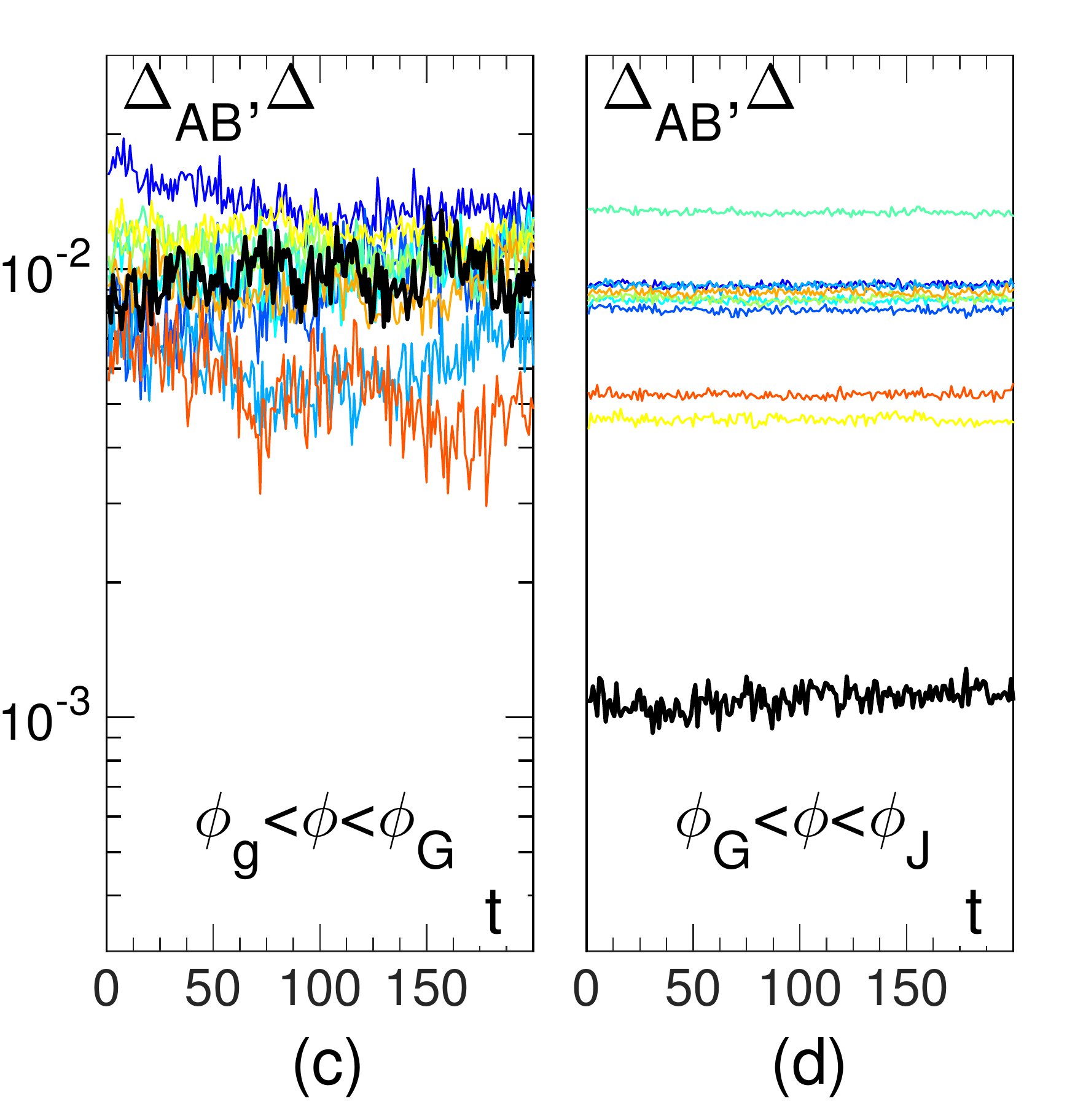}
\includegraphics[width=0.49\columnwidth,height=0.49\columnwidth]{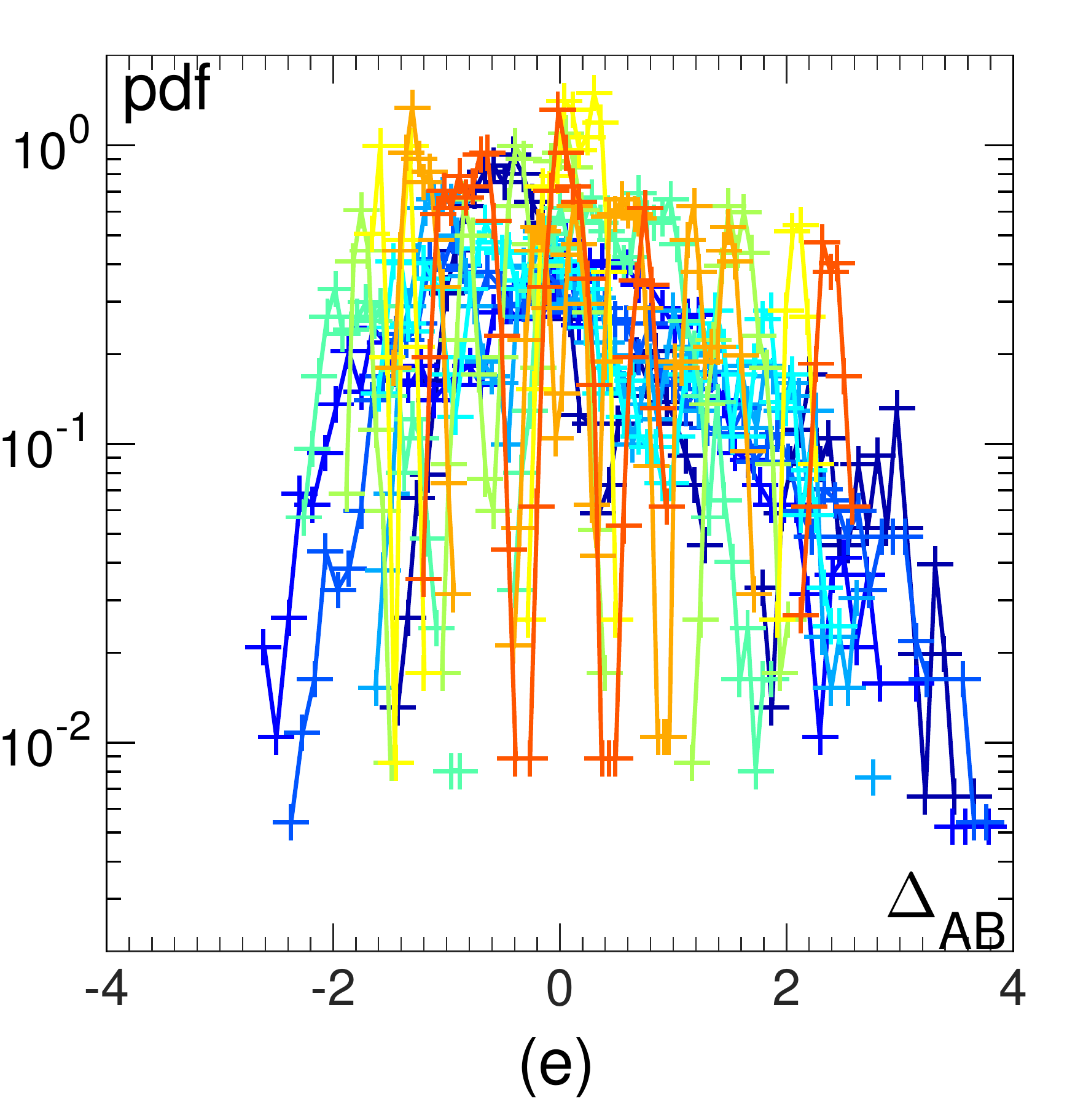}\\
\vspace{-0.3cm}
\caption{{\bf Statistics of $\Delta$ and $\Delta_{AB}$.} Probability distribution of $\Delta/<\Delta>$ sampled over time and compression cycles (a) or over time only (b); (inset : dynamical susceptibility associated to the fluctuations of $\Delta^k(t,\tau_0)$ over time). Temporal evolution of $\Delta^{k,k'}(t)$ for $10$ different pairs of compression cycles (color coded) for $\phi = 0.8196$ (c) and $\phi = 0.8228$ (d) as compared to the temporal evolution of $\Delta^k(t,\tau_0)$ (in black). Probability distribution function of $\Delta_{AB}$ sampled across compression cycles.}
\label{fig:pdfdeltas}
\vspace{-0.5cm}
\end{figure}
  
We now turn to the analysis of the fluctuations of $\Delta$ and $\Delta_{AB}$. Remember that we deal with a single glass sample. Hence fluctuations have only too possible sources: temporal fluctuations inside each compression cycle and inter-cycle fluctuations. The cage size fluctuations across time and cycles, normalized by the mean cage size, (fig.~\ref{fig:pdfdeltas}-a) present large exponential tails.  This observation strongly contrasts with that reported in the numerical study of the MK mean-field model~\cite{Charbonneau:2015ig}. In that case the distributions for $\Delta$ are Gaussian-like at all $\phi$. Fig.~\ref{fig:pdfdeltas}-(b) shows the distribution for the same quantity but sampled separately within each cycle. They exhibit similar exponential tails, which we thus attribute to temporal fluctuations of $\Delta^k(t,\tau_0)$. Such fluctuations highlight the presence of dynamical heterogeneities already reported for the same system~\cite{Lechenault:2008tw,Coulais:2012vj} as well as in numerical simulations of harmonic spheres~\cite{Ikeda:2013wp}. Those heterogeneities can be quantified by computing in each cycle, then averaging over the cycles, the associated dynamical susceptibly $\chi^4_{\Delta}(\tau_0) = {\rm N Var}\left(\Delta(t,\tau_0)\right/\sigma)$, where $\sigma$ is the standard deviation of the cage size for a single particle. This intra-state dynamical susceptibility presents a maximum around the Gardner transition (inset of fig.~\ref{fig:pdfdeltas}-b)). Although this susceptibility is not computed in~\cite{Berthier:2015db}, the distributions of $\Delta$, there also, present signs of non-gaussianity at $\phi_G$. This remarkable feature, absent from mean field results, is an interesting target for future theoretical developments in finite dimension.

The fluctuations of $\Delta_{AB}$ are better captured by following the temporal evolution of $\Delta^{k,k'}(t)$ for different pairs of compression cycles (fig.~\ref{fig:pdfdeltas}-c,d). When $\phi<\phi_G$, $\Delta^{k,k'}(t)$ fluctuates around the same value as $\Delta^k(t,\tau_0)$ and the fluctuations overlap from one compression cycle to another: the system explores the whole meta-basin of the glass. By contrast when $\phi>\phi_G$, the distances between different pairs of compressed states are very different and much larger than the typical cage size. In the numerical studies~\cite{Charbonneau:2015ig} and \cite{Berthier:2015db}, the authors report that the distributions for $\Delta_{AB}$ sampled over many samples, starting from a Gaussian-like distribution for $\phi<\phi_G$, develop exponential tails near $\phi_G$, before broadening again at larger $\phi$. As a result the skewness of the $\Delta_{AB}$ distributions exhibits a maximum around $\phi_G$. This skewness is attributed to sample to sample fluctuations, which are absent in the present experimental study. We only have access to the distributions of the reduce centered distribution of $\Delta_{AB}$ sampled over the pairs of cycles. Despite the strong lack of statistics (fig.~\ref{fig:pdfdeltas}-e), a systematic positive skewness is present here too. 

{\it --- Discussion and perspectives ---} 
We have conducted the first direct experimental observation of the Gardner phase, deep into the glass phase, before jamming. Although it has been obtained in a rather specific type of -- granular -- glass, there are now many evidences that such glasses are good models for hard potential thermal glass formers~\cite{Dauchot:2011vx}. At the level of the average order parameters of the transition, we obtain an excellent agreement with the results obtained in the numerical studies of the MK model~\cite{Charbonneau:2015ig} and hard sphere glasses~\cite{Berthier:2015db}. We observe some differences in the fluctuations statistics of the cage size; in particular our results suggest that intra-state dynamical heterogeneities are maximal at the Gardner transition. 
Further quantitative comparisons with existing numerical simulations require significantly more work, which are far beyond the scope of the present study, from both the numerical and experimental side. The importance of aging, the role of activated processes, and the finite size effects should be quantified precisely. To do so, studying the dynamics of the caging processes is obviously an important and promising next step.

The present study sheds new light on several results obtained previously in the same experimental set-up. 
In particular, it would be very interesting to investigate how the large dynamical heterogeneities observed just below jamming~\cite{Lechenault:2008tw,Lechenault:2010cl,Coulais:2012vj,Coulais:2014un}, are related to the marginality of the many glass states in the Gardner phase.
Another fascinating observation, which must contain a signature of the Gardner phase, is the avalanche dynamics observed in the motion of an intruder pulled through the glass, at constant force or constant velocity~\cite{Candelier:2009kb,Candelier:2010jl}. 
Finally, the non-linear elasticity and dilatancy reported in~\cite{Coulais:2014bn,Otsuki:2014hk,Tighe:2013ta}, might have close connections to the breakdown of classical elasticity reported for amorphous solid in the Gardner phase~\cite{Lerner:2013vf,Nakayama:2015wi,Dubey:2016bu,Biroli:2016th}.

{\it --- Acknowledgments ---}
We thank, Francesco Zamponi, Patrick Charboneau and Giulio Biroli for many interesting discussions related to this work and Yuliang Jin for assisting us in defining the compression protocols using preliminary simulations of our system. We also thank Vincent Padilla and C\'ecile Wiertel-Gasquet for invaluable technical support. Part of this work was supported by PSL grant JAM-T\&E and ÒInvestissements d'AvenirÓ LabEx PALM (ANR-10-LABX-0039-PALM).

\bibliography{/Users/olivierdauchot/Documents/_Science/Biblio/Glasses,/Users/olivierdauchot/Documents/_Science/Biblio/Jamming}


\end{document}